# Developing a Culturally Grounded, AI-Augmented UX Research Point of View (POV): An Exemplar Case Study from Telemedicine Dementia Care

Abiodun Adedeji
School of Computing and Informatics,
Bournemouth University Poole, UK,

Huseyin Dogan.
School of Computing and Engineering,
Bournemouth University Poole, UK,

Festus Adedoyin
School of Computing and Engineering,
Bournemouth University Poole, UK,

Michelle Heward
School of Computing and Engineering,
Bournemouth University Poole, UK,

Melike Akca
School of Computing and Engineering,
Bournemouth University Poole, UK,

Emmanuel Oluwatosin Oluokun
School of Computing and Engineering,
Bournemouth University Poole, UK,

Fatima Ahmad Muhazu
School of Computing and Engineering,
Bournemouth University Poole, UK,

Olumuyiwa Ayorinde
School of Computing and Engineering,
Bournemouth University Poole, UK,

User Experience Research (UXR) Points of View (POVs) distil complex and often fragmented research evidence into actionable perspectives that guide how teams interpret user needs, frame design decisions, and align stakeholders. Although POVs are widely used in industry practice, there are few published examples that explicitly document how POVs are constructed, particularly in culturally sensitive and low-resource contexts.
This paper presents an exemplar case study demonstrating how a culturally grounded, AI-augmented UXR POV was developed to inform TeleDeCa, a telemedicine dementia care framework for family caregivers in Nigeria.

Building on the UXR POV Playbook and pyramid framework, we illustrate how mixed-methods research, hypothesis generation, and ontology-based modelling can be combined to form a defensible POV without requiring a fully finalised system or validated outcomes.
Generative AI (GenAI) is integrated across the UXR POV framework as a bounded research collaborator, supporting synthesis, hypothesis exploration, and narrative construction while preserving human judgment, ethical accountability, and cultural sensitivity. The contribution of this paper lies in the extraction of reusable Play Cards and a Play that extend the UXR POV Playbook and serve as exemplar material for the CHI 2026 workshop on developing AI-powered UXR POVs.

CCS Concepts: Human-centered computing, User studies Human-centered computing, HCI theory, concepts and models Applied computing, Health care information systems Computing methodologies Artificial intelligence

**Additional Keywords and Phrases: User Experience Research; Point of View; UXR POV Playbook; Telemedicine; Dementia Care**

## 1 INTRODUCTION

User Experience Research (UXR) is increasingly expected to demonstrate strategic impact and cross-functional alignment [1,2]. Beyond generating findings, researchers must articulate defensible positions that integrate empirical evidence with contextual judgement. These demands intensify in healthcare, where cultural interpretation and infrastructural constraints directly shape design outcomes [3].
This study focuses on dementia caregiving in Nigeria, where care is predominantly informal and influenced by stigma, spiritual interpretations of illness, and uneven digital literacy [4]. In such contexts, usability alone does not ensure legitimacy or sustained engagement. Telemedicine systems must align with caregiving norms, infrastructural variability, and trust expectations. To support structured synthesis, we formalised an ontology of caregiver roles, care activities, telemedicine features, and contextual constraints, following established ontology design principles [5,6]. This semantic boundary enabled Generative AI (GenAI) to assist clustering and reframing while preserving researcher oversight.
A UXR Point of View (PoV) represents a defensible, evidence-based stance integrating data and contextual interpretation to guide design decisions [7]. Although widely used in practice, structured mechanisms for constructing and communicating PoVs remain limited, particularly in culturally sensitive settings [8]. The integration of GenAI further complicates PoV development. While AI can accelerate synthesis, it raises concerns regarding accountability, bias, and interpretive authority [8,9]. Human-AI interaction research emphasises meaningful oversight and clear governance boundaries [10,11].
Playbooks provide a structured means of externalising expertise under uncertainty [12,13]. In UXR, they help translate fragmented insight into defensible positions that support organisational alignment [1,7]. While collaboration challenges are well documented, systematic approaches for transforming heterogeneous evidence into PoVs remain underdeveloped. The UXR PoV Playbook addresses this gap through four stages Foundation, Data Collection, Insight Generation, and PoV articulation and reusable Cards and Plays [1]. Empirical demonstrations in culturally complex healthcare contexts are limited.
GenAI is increasingly incorporated into UX workflows for clustering and drafting [14]. However, outputs must remain provisional and subject to human validation [10,11]. Unconstrained systems risk amplifying bias and obscuring responsibility, and may affect interpretive ownership [15]. Consistent with sensemaking perspectives that treat interpretation as iterative and human-led, this study adopts a bounded GenAI approach [16,17].
Telemedicine extends healthcare access in resource-constrained settings, yet success depends on cultural alignment, infrastructural feasibility, and stakeholder trust [3]. In Sub-Saharan African dementia care, barriers include digital literacy variability, bandwidth instability, and culturally embedded illness interpretations [4]. These realities challenge UX approaches that prioritise usability metrics without contextual grounding.
Using TeleDeCa, a research-led telemedicine initiative supporting dementia caregivers in Nigeria [19], this paper demonstrates how the UXR PoV Playbook can be operationalised through ontology-constrained GenAI collaboration [1]. TeleDeCa is treated as a PoV-building artefact rather than a deployed system, illustrating governance-aware AI integration across the four PoV stages.
This paper contributes:

- An AI-augmented UXR PoV case study in a culturally sensitive telemedicine context.
- A structured demonstration of bounded GenAI integration across PoV stages.
- Reusable Playbook Cards derived through human-AI collaboration.
- A stakeholder-tailored PoV narrative aligned with the CHI 2026 workshop framework.

## 2 METHOD

This study employed an AI-augmented UXR synthesis using the UXR Point-of-View (PoV) framework [1,7]. TeleDeCa, a research-led telemedicine initiative supporting dementia caregivers in Nigeria, was used as a structured case to construct a culturally grounded PoV.

The method followed four stages: Foundation, Data consolidation, Insight and Play Card generation, and PoV articulation [7]. Generative AI supported clustering and reframing under researcher oversight. This approach combines mixed-methods synthesis with ontology-constrained AI reasoning to produce defensible PoVs for culturally sensitive healthcare contexts [20].

### 2.1 Data Sources and Evidence Identification

Multiple complementary sources were synthesised, including a systematic review of telemedicine and cultural competency in dementia care in Sub-Saharan Africa [19], a scoping review of telemedicine adoption in Nigeria [4], semi-structured interviews with caregivers and healthcare professionals, a survey assessing caregiver burden and digital literacy, and participatory design workshops.

Qualitative data were analysed using thematic analysis with survey findings examined descriptively [20]. The dataset spanned cultural, emotional, infrastructural, and technological dimensions, an ontology-based model was developed to preserve coherence. Using Resource Description Framework concepts, relationships among caregiver roles, care activities, emotional states, telemedicine features, and contextual constraints were specified in figure 1. The ontology constrained Generative AI abstraction and reduced decontextualized generalization [5,6].

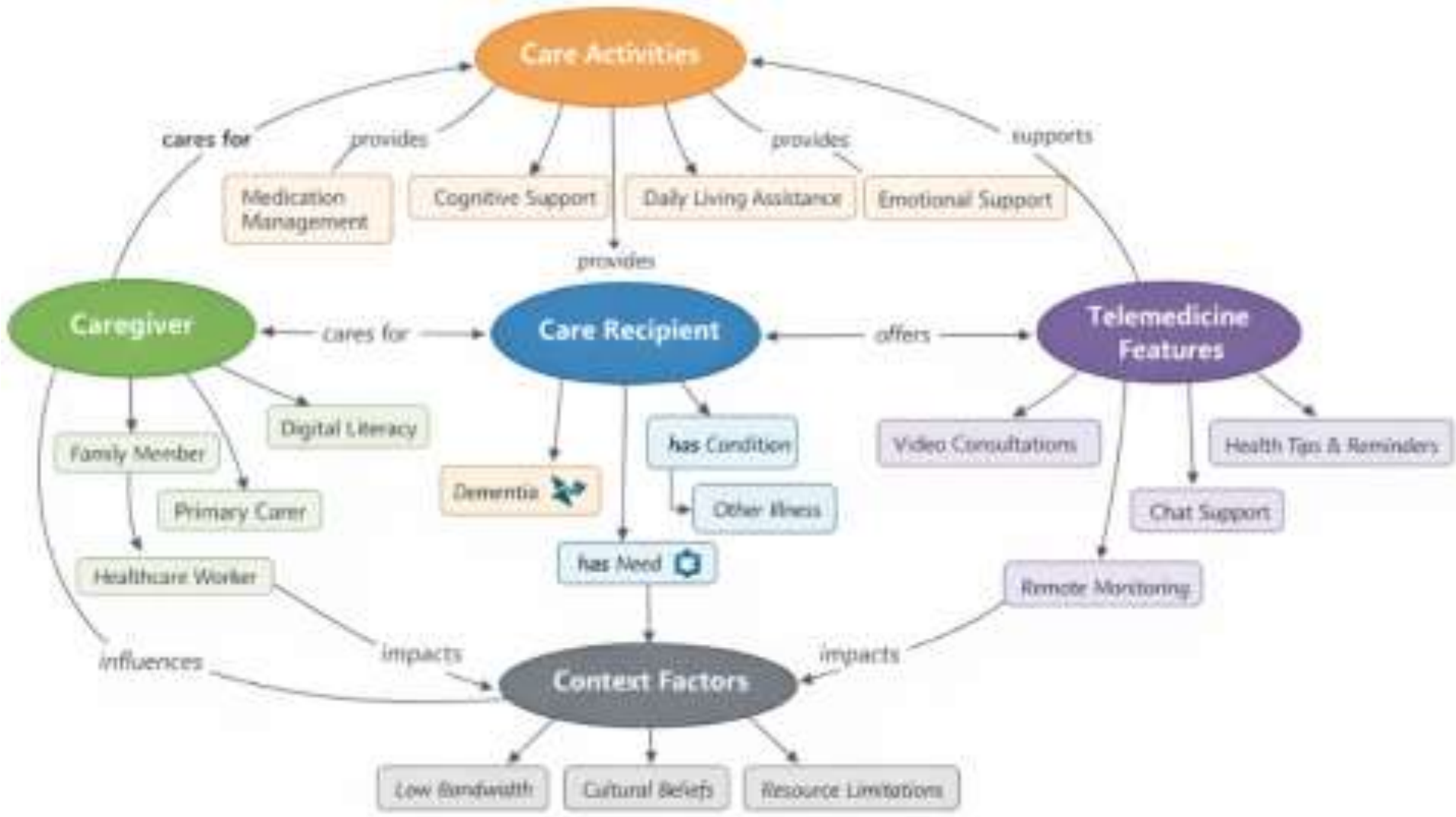

Figure 1. Ontology-based boundary model used to constrain GenAI abstraction across caregiver roles, care activities, telemedicine features, and contextual factors in the TeleDeCa case study.

### 2.2 Procedure and Prompting Strategy

This study operationalised the four-stage AI-powered UXR PoV framework within the TeleDeCa case study. Generative AI was used as a bounded analytical collaborator to support structuring, clustering, and reframing. Interpretive authority remained with the research team throughout.

**Stage 1: GenAI-Supported Evidence Structuring**

This stage consolidated heterogeneous evidence into design-relevant clusters. Structured summaries from reviews, interviews, and surveys were uploaded to GenAI to identify recurring determinants shaping telemedicine feasibility in Nigerian dementia care, not to generate hypotheses. GenAI grouped themes around emotional burden, stigma, digital literacy variability, infrastructural instability, and trust fragility, aligned with the UXR PoV progression (Foundation → Data → Insight). These validated clusters informed subsequent stakeholder mapping.

Prompts included:

- "Analyse the structured TeleDeCa evidence corpus and identify recurring caregiver, infrastructural, and trust-related challenges influencing telemedicine adoption."
- "Cluster these themes into contextual, behavioural, and design-level determinants aligned with the UXR PoV framework."
- "Summarise stable patterns that may shape telemedicine system design."

Outputs at this stage consisted of validated thematic clusters, not predictive claims.

**Stage 2: Foundational Planning and Stakeholder Road mapping**

The second stage translated thematic clusters into a structured stakeholder matrix. GenAI was prompted to map interdependencies across caregivers, healthcare professionals, community actors, policymakers, UX designers, UX researchers, and software developers.

This stage clarified systemic tensions, including balancing emotional reassurance with technical feasibility, and cultural grounding with scalable design logic. Stakeholder roles, core needs, constraints, and resulting design implications were formalised.

Prompts included:

- "Based on the TeleDeCa clusters, explain how caregivers experience emotional burden and digital barriers in dementia telemedicine contexts."
- "Identify primary and secondary stakeholders including caregivers, clinicians, UX designers, researchers, and developers. Describe their roles, constraints, and decision authority."
- "Map tensions between emotional safety, infrastructural realism, and scalable implementation."

Stage 2 established the structural conditions required for design translation.

**Stage 3: Insight Generation and Playbook Card Development**

Stage 3 translated recurring tensions into structured UXR Playbook Cards using the Building Blocks model (Foundation → Data → Insight → PoV translation). GenAI supported articulation of issue statements and preliminary best-practice guidance. Outputs were subsequently refined manually to align with the Playbook structure (Issue → Best Practice → Design Implication → Related Cards) and ensure contextual accuracy.

Four exemplar Playbook Cards emerged (Figure 3):

- P1- Cultural Grounding as a Precondition for Usability,
- P2- Infrastructural Constraint as Primary Design Boundary,
- P3- Emotional Load as a Core Design Variable, and
- P4- Explainability as Trust Infrastructure.

These cards formalised contextual determinants into reusable design artefacts.

Prompts included:

- Translate the identified TeleDeCa determinants and stakeholder tensions into structured UXR Playbook Cards."
- "For each issue, articulate the contextual evidence and propose a best-practice design response grounded in the synthesised findings."
- "Ensure alignment with cultural grounding, emotional reassurance, infrastructural realism, and explainable AI principles."
- "Use the UXR PoV Playbook Card template to structure the issue statement and corresponding guidance."

**Stage 4: PoV Narrative Construction and Stakeholder Communication**

The final stage translated Playbook insights into stakeholder-specific PoV narratives tailored to caregivers, healthcare professionals, UX designers, UX researchers, policymakers, and software developers. GenAI supported narrative condensation and role-specific framing. All PoV statements were validated and refined to preserve cultural specificity and implementation realism.

Prompts included:

- "Generate stakeholder-specific PoV narratives for caregivers, clinicians, UX designers, researchers, policymakers, and developers."
- "Refine each PoV to align with emotional reassurance, infrastructural realism, and explainability requirements."
- "Ensure implementation feasibility is explicitly addressed for software developers."

Together, these four stages demonstrate a governance-aware integration of GenAI within the UXR PoV framework, progressing from contextual grounding to stakeholder-aligned design articulation.

## 3 RESULTS

This study demonstrates how Generative AI (GenAI), used as a bounded synthesis collaborator, supports development of a culturally grounded User Experience Research (UXR) Point of View (PoV) for telemedicine dementia care in Nigeria. Findings are organised across the four-stage AI-powered UXR PoV framework: (1) contextual synthesis, (2) stakeholder road mapping, (3) Playbook Card development, and (4) stakeholder-aligned PoV articulation.

Empirical material from reviews, interviews, surveys, and workshops was progressively structured into design-relevant guidance. Multiple GenAI systems were used for clustering and reframing; despite minor stylistic differences, outputs converged across themes and stakeholder mappings, indicating stable patterns within the TeleDeCa evidence base.

### 3.1 Leveraging GenAI for Contextual Synthesis

In the first phase, Generative AI (ChatGPT-5.2) was used as a bounded synthesis partner to consolidate TeleDeCa evidence from reviews, interviews, surveys, and workshops. Through ontology-constrained prompting and researcher validation, recurring determinants of telemedicine feasibility were identified within the UXR PoV progression.
Four determinants emerged:

- Emotional burden shaping caregiver interaction
- Digital literacy gradients constraining access
- Trust fragility toward opaque AI systems
- Stigma influencing dementia interpretation and adoption

Engagement was driven less by features than by cultural legitimacy, reassurance, and infrastructural realism. These determinants informed subsequent stakeholder mapping, Playbook Cards, and PoV development. As illustrated in Figure 2, these determinants were mapped across the four stages of the UXR PoV framework, progressing from contextual grounding to articulated narrative.

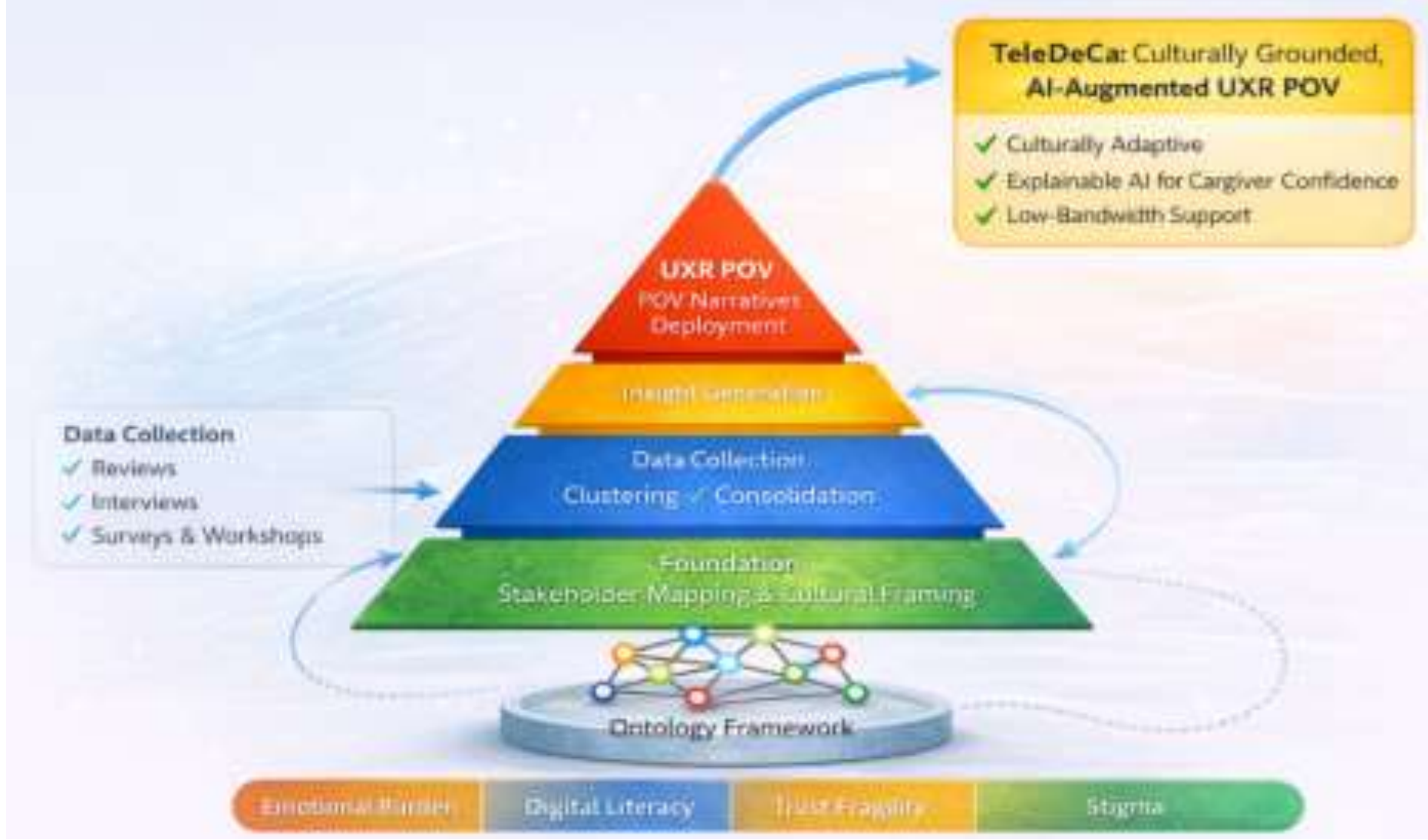

Figure 2. Mapping of TeleDeCa findings to the four stages of the UXR PoV framework, illustrating progression from contextual grounding to PoV articulation.

### 3.2 Establishing a Stakeholder Roadmap

In the second stage, Generative AI (ChatGPT-5.2) synthesised stakeholder perspectives shaping TeleDeCa telemedicine design. A structured roadmap identified four primary groups: family caregivers, healthcare professionals, community actors, and policymakers.
Caregivers require emotionally safe, low-complexity systems but face cognitive strain and digital variability. Healthcare professionals need progress visibility within existing workflows. Community actors shape trust and adoption norms. Policymakers determine infrastructural feasibility and scalability.
This mapping clarified interdependencies and key tensions, particularly balancing emotional safety with technical feasibility and cultural grounding with scalable implementation.

### 3.3 TeleDeCa UXR Playbook Cards

In the third stage, validated determinants and stakeholder tensions were translated into a structured set of TeleDeCa UXR Playbook Cards. Developed through iterative GenAI-assisted drafting (ChatGPT-5.2) and researcher validation, these cards operationalise contextual evidence into actionable guidance for designers, researchers, and decision-makers working in culturally sensitive telemedicine contexts.
Each card captures a recurring design tension and formalises four elements: a clearly defined issue, contextual grounding in the TeleDeCa evidence base, a best-practice design response, and links to related themes within the broader PoV framework. This structure enables translation from cultural and infrastructural insight to concrete design direction.
Four exemplar Playbook Cards emerged (Figure 3):

- P1- Cultural Grounding as a Precondition for Usability, positioning cultural legitimacy as foundational to interface effectiveness.

- P2- Infrastructural Constraint as Primary Design Boundary, framing bandwidth and device variability as core design parameters.
- P3- Emotional Load as a Core Design Variable, prioritising reassurance and cognitive simplicity over feature expansion.
- P4- Explainability as Trust Infrastructure, treating transparency as essential for sustained caregiver engagement.

Together, these cards establish a shared, evidence-grounded design language linking contextual reality to strategic UX decisions.

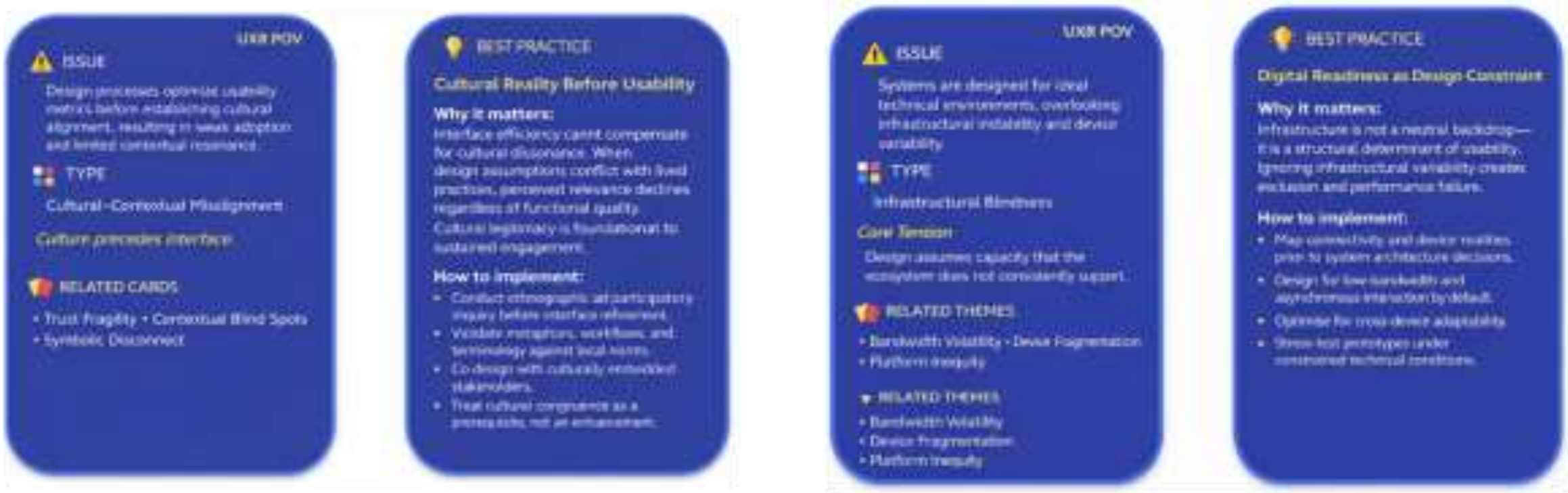

*P1- Cultural Grounding as a Precondition for Usability* *P2- Infrastructural Constraint as Primary*

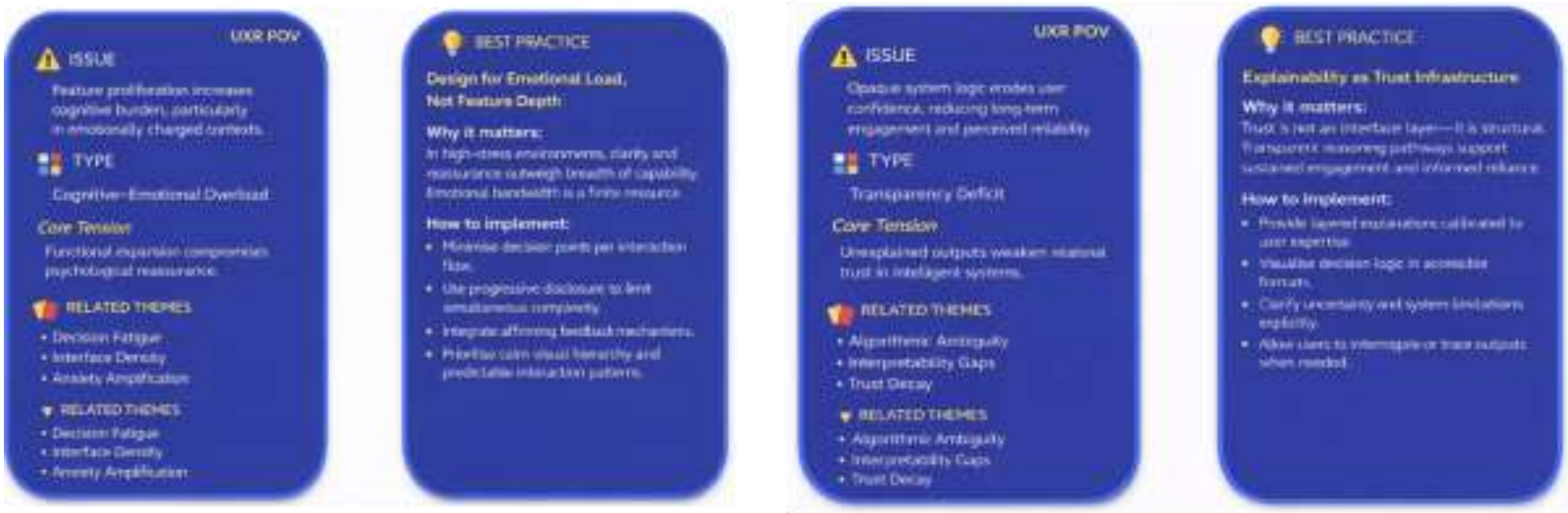

*P3- Emotional Load as a Core Design Variable* *P4- Explainability as Trust*

Figure 3. Exemplar TeleDeCa Playbook Cards demonstrating structured issue framing, risk articulation, and reusable guidance.

### 3.4 Stakeholder-Aligned PoV Narratives

In the final stage, stakeholder-specific PoV narratives translated TeleDeCa Playbook insights into strategic guidance. Generative AI (ChatGPT-5.2) supported phrasing and condensation within a researcher-led process.

For caregivers, the PoV prioritised emotionally supportive, low-bandwidth systems. Healthcare professionals required interpretable dashboards and explainable AI aligned with workflows. Policymakers emphasised scalability and infrastructural realism, while community actors focused on trust-centred communication.

Across groups, four pillars emerged: cultural grounding, emotional reassurance, infrastructural realism, and explainable AI. In figure 4, a cross-stage "Stakeholder Buy-In" play integrated multiple cards to address organisational resistance while preserving cultural integrity.

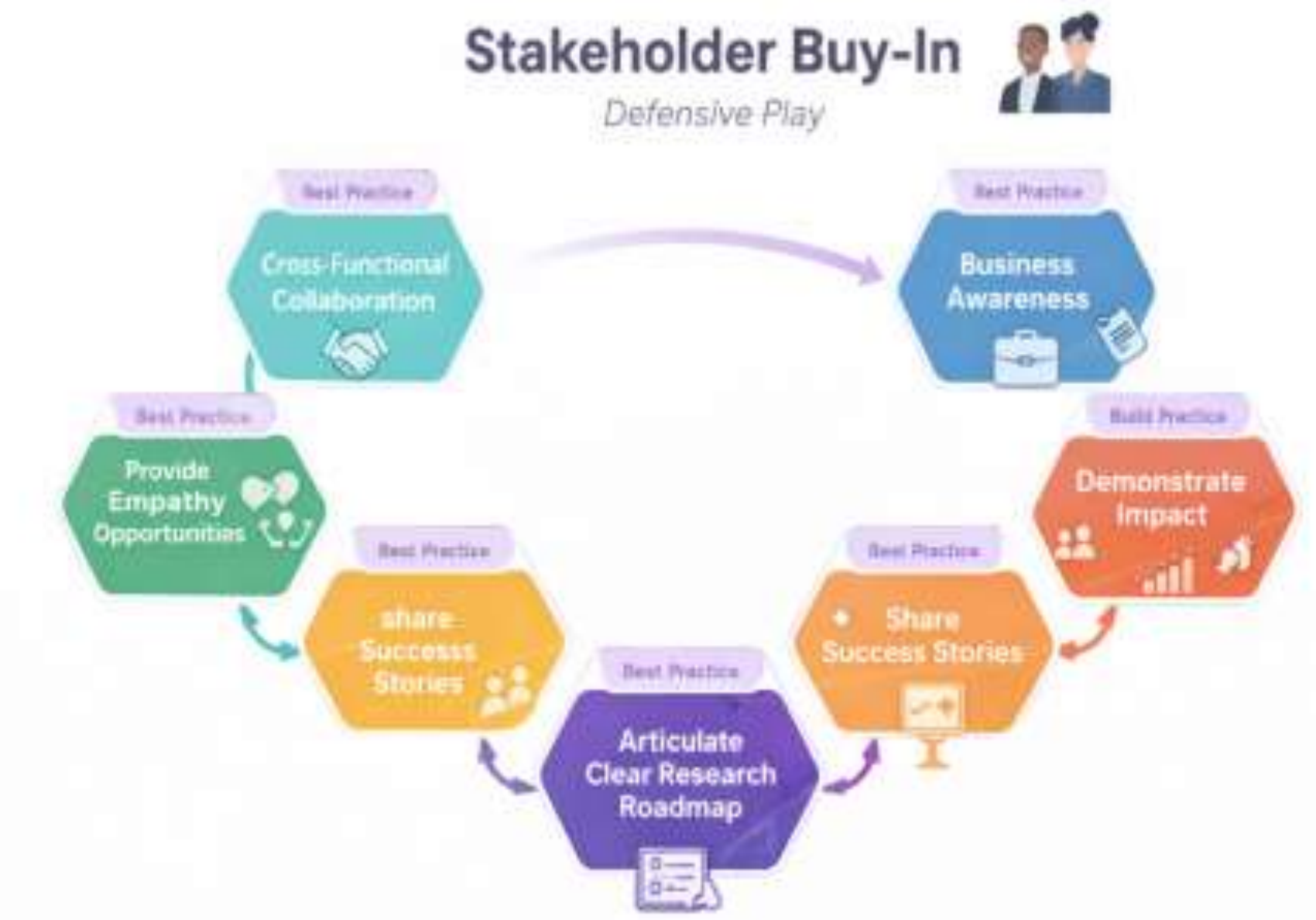

Figure 4. Stakeholder Buy-In defensive play illustrating cross-stage integration of Playbook Cards.

## 4 DISCUSSIONS

This study operationalised the UXR PoV Playbook [1,7] as an AI-augmented synthesis methodology within a culturally grounded telemedicine context. Using TeleDeCa [4,19], it demonstrated how ontology-constrained Generative AI (GenAI) can scaffold progression across the four PoV stages while preserving human interpretive control. GenAI improved structural organisation and thematic clustering but introduced epistemic risks requiring mitigation. GenAI functioned strictly as a bounded structuring tool; interpretive claims and contextual judgements remained exclusively researcher-authored

GenAI consolidated interviews, surveys, and review findings into stable themes, emotional burden, digital literacy variability, trust fragility, and stigma supporting emerging evidence that GenAI tools can assist UX research synthesis and workflow acceleration [14,2]. However, it lacked sensitivity to cultural nuance. Dementia care in Nigeria is shaped by stigma and informal caregiving norms [19], which cannot be inferred from pattern regularities alone. Consistent with established human-AI interaction guidelines, all AI outputs were systematically validated against primary empirical material [19].

Generative systems may also obscure accountability and reshape interpretive ownership during collaborative analysis [8,15]. Ontology-based constraints grounded in formal knowledge representation principles reduced abstraction drift, yet reflexive oversight remained necessary. Cross-model comparison revealed thematic convergence with minor stylistic variation, consistent with documented concerns regarding epistemic risk and alignment in language model deployment [5,6].

Methodologically, GenAI reduced cognitive load during synthesis and facilitated translation of contextual insights into reusable Playbook Cards [1]. However, responsible integration requires boundary constraints, triangulation, and explicit human validation. The study therefore advances a governance-aware application of the UXR PoV framework, positioning GenAI as a bounded collaborator rather than an autonomous analytic agent embedded within accountable research practice [16,11].

## 5 CONCLUSIONS

This paper presents an AI-augmented operationalisation of the UXR PoV Playbook within a culturally sensitive telemedicine case study. By constraining GenAI to structured synthesis and reframing roles across Foundation, Data consolidation, Insight Generation, and PoV articulation, the study shows how AI can enhance analytical efficiency without displacing human accountability. The core contribution is a governance-aware demonstration of AI-assisted UXR PoV construction in a culturally grounded healthcare context.

Future work should examine cross-domain validation, longitudinal evaluation in live telemedicine deployments, and systematic assessment of ontology-based constraint mechanisms. Further research is also needed to investigate the cognitive and motivational implications of sustained human-AI collaboration in research synthesis workflows.